\documentclass[9pt,twocolumn,twoside]{pnas-new}
\usepackage{bm}
\usepackage{bbm}
\usepackage{caption}
\usepackage{subcaption}
\templatetype{pnasresearcharticle}

\title{Self-organised criticality in high frequency finance: \\
the case of flash crashes}

\author[a,1]{Jeremy D. Turiel}
\author[a,b]{Tomaso Aste} 

\affil[a]{Department of Computer Science, UCL, Gower Street, WC1E6BT London, UK}
\affil[b]{Systemic Risk Centre, London School of Economics and Political Sciences, London, United Kingdom}
\affil[1]{Corresponding author. E-mail: jeremy.turiel@gmail.com}

\leadauthor{Turiel}

\authorcontributions{Please provide details of author contributions here.}
\authordeclaration{Please declare any conflict of interest here.}
\equalauthors{\textsuperscript{1}A.O.(Author One) and A.T. (Author Two) contributed equally to this work (remove if not applicable).}

\begin{abstract}
With the rise of computing and artificial intelligence, advanced modeling and forecasting has been applied to High Frequency markets. A crucial element of solid production modeling though relies on the investigation of data distributions and how they relate to modeling assumptions. In this work we investigate volume distributions during anomalous price events and show how their tail exponents $<2$ indicate a diverging second moment of the distribution, i.e. variance. We then tie the dynamics of flash crashes to self-organised criticality. The findings are of great relevance for regulators and market makers as they advocate for rigorous heavy-tailed modeling of risk and changes in regulation to avoid simultaneous liquidity withdrawals and hard risk constraints which lead to synchronisation and critical events.
\end{abstract}

\dates{\today}

\begin{document}

\maketitle

\ifthenelse{\boolean{shortarticle}}{\ifthenelse{\boolean{singlecolumn}}{\abscontentformatted}{\abscontent}}{}

\section{\label{intro}Introduction}

High frequency trading and the dynamics of market microstructure have seen growing interest in both academia and industry, with the rise of electronic trading and the emergence of major HFT market makers in recent years. Electronic trading has led to an increase in trading volumes and frequencies particularly in the last decade, which sparked an investigation into how market dynamics have evolved with market players.

The literature and practitioners alike have noted that price efficiency has improved as a result, but also that exchanges, assets and even markets now carry more systemic risk as a result of large players covering the whole space with similar methods and risk constraints shared across assets and exchanges. The growth in systemic risk due to electronic trading has given rise to a growing number of increasingly large flash crashes \cite{calcagnile2018collective} of which the one of May 6th 2010 was the first notorious example.

The report on the May 6th crash by Nanex \cite{nanex2010nanex} begins to suggest how high frequency order placement and saturation might have worsened the extent of the crash. The authors in \cite{johnson2012financial} delve deeper into this idea and investigate flash crashes between 2006 and 2011 to show a system-wide phase transition around $\sim 500ms$ to an all-electronic trading market characterised by black swan events. Further insight into the impact of HFT players in flash crashes was provided by the simulations in \cite{paddrik2012agent}, where the authors show how lowering the number of HFT players in the simulation reduces the extent of the crash, even when the size of the large sell order which triggered it is kept constant.

The authors explain the relation between the number of HFT players and the extent of the crash as due to the ``hot potato'' phenomenon described in \cite{golub2012high}. This phenomenon can be explained as follows. When an unusually large sell (buy) order hits the market it gets absorbed by liquidity providers (often HFT market makers) which, as a result, accumulate a short position in their inventory. The unusually large order though impacts the price and potentially triggers risk limits by those market makers holding the inventory. Meanwhile as the price drops (rises) dramatically ordinary market players withdraw from the market.
As a result of the risk limits HFTs try to reduce their inventory with aggressive market orders. As everyone else has withdrawn, they trade with each other at extremely high frequency, thereby creating high trading volumes. This particular phase characterises the ``hot potato'' phenomenon, i.e. the inventory exposure being passed around like a ``hot potato''.

The positive feedback loop continues as follows. When trading volume is used as a proxy for liquidity the ``hot potato'' phenomenon creates apparent liquidity which triggers execution by lower frequency players which are also trying to reduce exposure as the market drops.

The authors in \cite{golub2012high} make two further points on HFTs and modern market dynamics during these extreme events. Another cause of apparent liquidity and market depth is fleeing liquidity by HFT market makers. The fast cancellation of limit orders has been the extensive topic of discussion and was suggested to create the illusion of market depth with the consequences discussed above.

As the number of trading venues and exchanges rises liquidity is fragmented. Sweep orders and arbitrageurs aim to move liquidity and remove arbitrage opportunities across exchanges, but can create dangerous effects.
Sweep orders between markets operate at lower frequencies than these fast cancellations, thereby sweeping the book after liquidity has potentially already been removed. This worsens the systemic aspect of these events across the fragmented liquidity structure of exchanges.

Extreme events of this kind, which are characterised by non-linear reactions to shocks in the system and positive feedback loops, exist in man-made and natural systems and are instances of self-organised criticality. In this type of systemic events the system reaches a critical state where a small release in energy or imbalance triggers highly non-linear reactions in size. This is the case of avalanches and more, as described in \cite{christensen2005complexity}.

Events with such underlying dynamics are characterised by heavy-tailed and in particular power law distributions. The tail's decay exponent in these distributions is crucial for their modeling as it indicates which moments of the distribution are defined and which diverge. This can be of great practical relevance for both regulators and practitioners, in particular market makers. These are market players with an obligation to provide liquidity at all times, such as banks, which cannot simply withdraw from the market under extreme conditions, but must continue to provide prices for their clients to trade at. This constraint brings the need to model trade volume flow in the market as well as from clients also under extreme market conditions. Predictive tools for volume modeling which rely on advanced statistical methods and Artificial Intelligence are now being adopted by most market makers, but the underlying data distribution of the phenomenon is crucial to estimate confidence intervals on these models. For example, unbounded distributions have major implications for the use of Gaussian-based versus distribution-agnostic (quantiles) confidence intervals and how extreme values are modeled and accounted for in terms of pricing and inventory management. Our work makes the argument that power law distributions and critical dynamics must be included in modern modeling, in particular when dealing with extreme events is the focus, as in risk modeling.

From the above discussion we have shown the relevance and interest in these extreme events and how their origin was investigated in particular through simulations.
Despite the great relevance of these properties, to our knowledge no investigation of the statistical properties of these events in their microstructural dynamics has been conducted yet. The statistics underlying extreme events are of great relevance for the study of the dynamics of these phenomena, but mostly for market makers who aim to provide liquidity and stability during such events and should have the tools to model and best navigate them.

In this work we bridge this gap by investigating flash crash events in high frequency markets from the perspective of critical systems and avalanche-like dynamics to show how the detected flash crashes constitute black swan events with unbounded trading volume distributions while the remaining order flow is sub-critical and bounded. This is of great importance for the modeling of order flow (and related assumptions) in both normal and extreme market conditions.

\section{\label{data}Data}

In the present work we consider granular order flow data for Apple (AAPL) from the NASDAQ exchange between 3/1/2017 and 25/9/2020. High frequency price data is obtained from LOBSTER \cite{huang2011} and sampled to obtain non-overlapping one minute returns. This frequency was also adopted in \cite{calcagnile2018collective} and other works in the literature for the detection of price jumps as it is understood that below this limit microstructural noise becomes relevant and can impact the validity of the method.
The data from LOBSTER provides detailed order flow data which indicates the type (limit order, cancellation, execution of a limit order), the side (buy or sell) and volume of each event in the order flow. This allows to isolate trading volume and other quantities from the rest of the order flow and we discuss the importance of this in Section \ref{method}.

\section{\label{method}Method}

In the present work we analyse the distribution of aggregate volumes by minute for each event type (independently of the side). We then detect price jumps and compare those distributions between intervals characterised by jumps and ones which are not.
We focus particularly on trades based on findings from the literature on market microstructure which has identified them as the most informative type of market event in the order flow \cite{potters2003more}, upon which much of the market impact literature is based \cite{bouchaud2009markets, bouchaud2004fluctuations, bouchaud2006random}.

\subsection{Jump Detection}\label{sec:method_jumps}

In the present work we detect price jumps (up and down crashes) similarly to \cite{calcagnile2018collective}, in 1 minute non-overlapping returns.

We apply the basic jump detection method from \cite{lee2008jumps} and detect jumps at the $1\%$ significance level. In addition to the basic features of the method for robust volatility estimation we obtain an estimate of intraweek periodicity and adjust the return series and jump detection according to \cite{boudt2011robust} as well.

\subsection{Power law fit}\label{sec:method_powerlaw}

We analyse the volume distributions for the 1min intervals and fit power law functions to the tails of the empirical Complementary Cumulative Distribution Function (CCDF). For fitting purposes, we define the tail as the percentile interval $[90\%, 99.9\%]$, where the highest values are excluded to reduce fitting noise and discount finite size effects. We highlight that the results are robust to different choices of tail quantile intervals, as will be discussed in Section \ref{results}.
The fit is then simply applied as a linear fit (minimum squared error) to the log-log data.


\section{\label{results}Results}

\begin{figure}[h]
     \centering
     \begin{subfigure}[b]{1\linewidth}
         \centering
         \includegraphics[width=\linewidth]{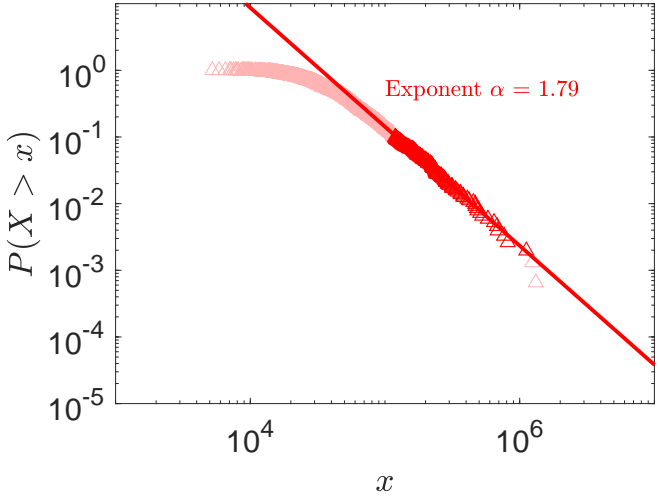}
         \caption{Volume distribution and fit for price jump intervals. We find a decay exponent $<2$ which indicates an unbounded critical distribution.}
         \label{fig:crash_volume_distributions}
     \end{subfigure}
     \hfill
     \begin{subfigure}[b]{1\linewidth}
         \centering
         \includegraphics[width=\linewidth]{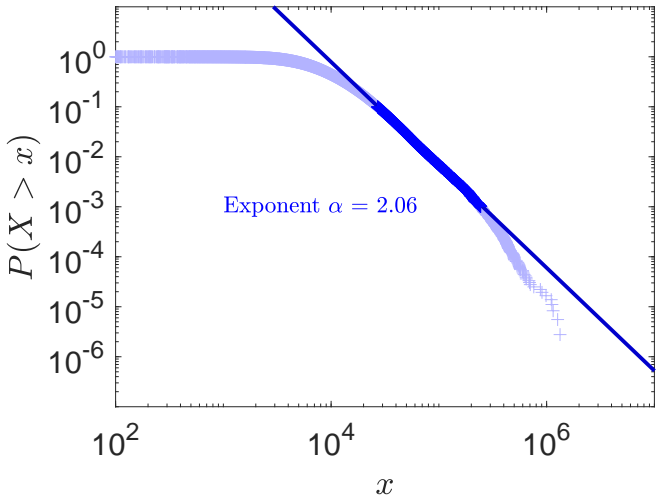}
         \caption{Volume distribution and fit for non-jump intervals. We find a decay exponent $>3$ which indicates a bounded distribution with well defined mean and variance.}
         \label{fig:normal_volume_distributions}
     \end{subfigure}
        \caption{{\bf Trading volume distributions} Log-log plot of the volume distributions for 1 minute intervals. The plots report the empirical distribution and power law tail fit for intervals with jumps (Figure \textbf{a}) and no jumps (Figure \textbf{b}).}
        \label{fig:volume_distributions}
\end{figure}

The plots in Figure \ref{fig:volume_distributions} show the distribution of aggregate trade volumes by minute for time intervals when a price jump was detected (Figure \ref{fig:crash_volume_distributions}) and those when it was not (Figure \ref{fig:normal_volume_distributions}).

We indeed observe the distribution of jump volumes to be shifted in mean and median and characterised by higher volumes, but this was expected from the relation between traded volume and price impact \cite{potters2003more}. What is interesting is the fact that the jump distribution is not only shifted with respect to the general distribution, but also characterised by a heavier tail. At this point we highlight that our jump detection method described in Section \ref{sec:method_jumps} is only aware of price changes, thereby removing any bias on explicitly selecting anomalous volumes, apart from their potentially anomalous corresponding price jump.

We apply the methodology from Section \ref{sec:method_powerlaw} to fit the power law tails of the jump and non-jump trade volume distribution CCDFs (Complementary Cumulative Distribution Function). The decay exponents are reported in Figure \ref{fig:volume_distributions}. We notice how the decay exponent of jump volumes is $<2$ in modulus, while that of non-jump volumes is $>2$ in modulus.
Decay exponents in power law distributions are of great importance \cite{christensen2005complexity} as they indicate which moments of the distribution are defined. Indeed a decay exponent $<2$ of the probability density function indicates that the second moment of the distribution (variance) is not defined, i.e. the distribution is unbounded, while one $>2$ characterises a bounded distribution with well-defined variance. This can also be viewed as the fact that the sample variance diverges with sample size for exponent $<2$, while it converges to the ``true'' variance in the limit, for exponent $>2$.

As the exponent values are crucial for our discussion, we wish to clarify some robustness properties of the results in relation to our description of the fitting methodology in Section \ref{sec:method_powerlaw}.

We choose the percentile interval $[90\%, 99.9\%]$ to fit the power law tail, but our main results are robust to choosing a starting level of say $95\%$ and not excluding the last $99.9\%$. The latter in particular gives a non-jump exponent $>3$ ($\alpha = 3.21$) and a virtually unchanged jump exponent. This shows how the exponent in Figure \ref{fig:normal_volume_distributions} is a lower bound and our qualitative conclusions are robust.

Another potential critique could arise from the fact that our fit in Figure \ref{fig:normal_volume_distributions} excludes jump intervals and this could be the cause of the observed bounded variance as extreme values are removed. We have indeed investigated the more comprehensive distribution which includes both jump and non-jump intervals and found that this has little statistical effect on the fit and resulting exponent ($\alpha = 2.02$). This also highlights that our filtering of jump intervals alone shows the unboundedness in their volume distribution and further supports our analysis and the validity of its conclusions.

A very important consequence of the above observations of the different decay exponent between jump and non-jump distributions, rather than just a shift, is the fact that indeed the two sets of intervals have different underlying distributions, generating processes and resulting characteristics. We report that the idea that the order flow distribution is not fully stationary and that it might influence the impact and subsequent order flow is already hinted by Bouchaud et al. in \cite{eisler2012price} on the basis of previous studies on impact \cite{farmer2006market, gerig2008theory, bouchaud2009markets}.

\begin{figure}[h]
     \centering
     \begin{subfigure}[b]{1\linewidth}
         \centering
         \includegraphics[width=\linewidth]{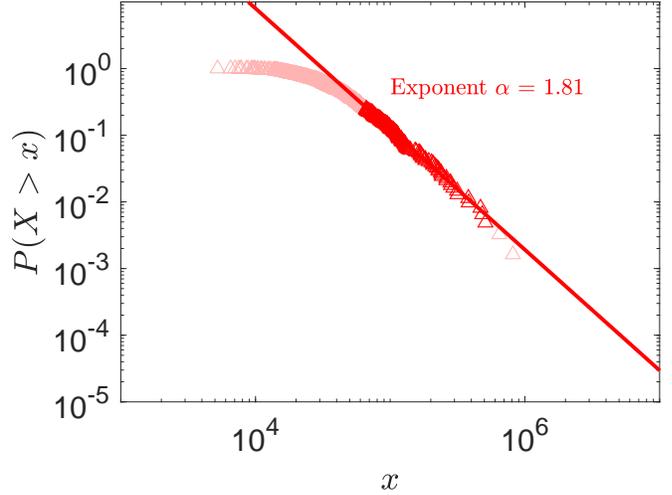}
         \caption{Volume distribution and fit for price jump intervals with \textbf{positive} return. We find a decay exponent $<2$ which indicates an unbounded critical distribution.}
         \label{fig:crash_volume_distributions_positive_return}
     \end{subfigure}
     \hfill
     \begin{subfigure}[b]{1\linewidth}
         \centering
         \includegraphics[width=\linewidth]{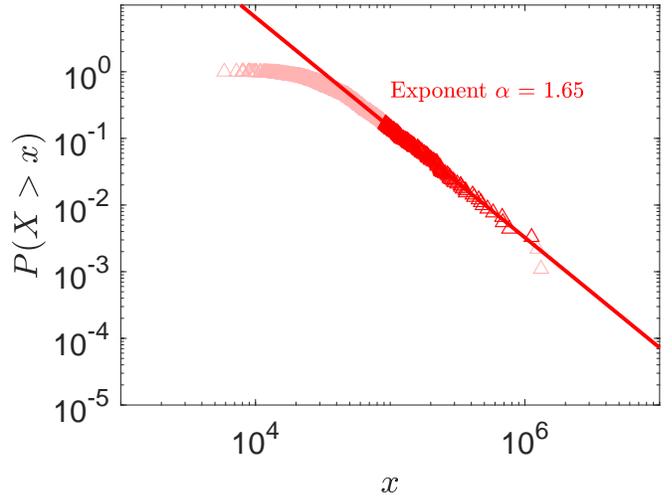}
         \caption{Volume distribution and fit for price jump intervals with \textbf{negative} return. We find a decay exponent $<2$ which indicates an unbounded critical distribution.}
         \label{fig:crash_volume_distributions_negative_return}
     \end{subfigure}
        \caption{{\bf Trading volume distributions by return sign} Log-log plot of the volume distributions for 1 minute intervals. The plots report the empirical distribution and power law tail fit for intervals with jumps and positive (Figure \textbf{a}) and negative (Figure \textbf{b}) corresponding price returns.}
        \label{fig:volume_distributions_returns}
\end{figure}

Our finding adds the essential point to the microstructural literature than indeed volume distributions are heavy-tailed, but that under ``normal'' market conditions the distribution is bounded and can be well approximated by a normal distribution with well-defined variance. On the other hand, black swan events in volume bursts belong to a different distribution and process altogether.
From the perspective of self-organised criticality, anomalous price jumps present an underlying volume process characterised by black swan events of onbounded magnitude where events of arbitrarily large sizes are characterised by non-zero probabilities.

The distinction is crucial in this context as black swan events characterise unbounded distributions alone and we observe how these extreme and anomalous price events are characterised by black swan events in the underlying market activity. This indeed supports the thesis from the literature described in Section \ref{intro} that crashes are characterised by a very large and concentrated trade activity with unbounded variance, different from the trade distribution of regular order flow.

A generalisation of the results from Figure \ref{fig:crash_volume_distributions} is presented in Figure \ref{fig:volume_distributions_returns}, which provides further insights into the dynamics of crashes in relation to known stylised facts.

It is broadly accepted in the literature that the return distribution is asymmetric with a heavier tail for negative returns \cite{bouchaud2018trades}. Based on the relation between volume imbalance and price returns \cite{cont2014price} one could suggest that the detected jumps correspond to negative returns which present a heavier tailed volume distribution. We show that our jump detection method accounts for this asymmetry as we detect a similar number of positive and negative jumps. Further, exponents for the volume distributions of positive and negative return intervals in Figure \ref{fig:volume_distributions_returns} are both consistent with our previous findings from Figure \ref{fig:volume_distributions}. However, we highlight that, as expected, negative return intervals are characterised by a heavier distribution tail as events are more extreme to the downside.

All the above findings have crucial practical implications for financial modeling where Normality assumptions are still widespread, in particular in risk models. We highlight that extreme events originating from a heavy-tailed distribution with unbounded variance are not compatible with assumptions of Normality and their unbounded variance causes to overestimate significantly the variance of the ``normal'' underlying volume distribution.

The particular features of the trade volume distribution in flash crashes support the anecdotal explanation of the ``hot potato'' phenomenon in the literature \cite{golub2012high} which suggests that HFT market makers stay in the market under extreme conditions and absorb large trades which cause them to build up large inventories. Once inventory limits are hit the players look to reduce it more aggressively as the price moves against them, but can only trade with each other. This causes the positive feedback loop of frenetic trading activities between these market makers trying to reduce their exposure. On top of the market makers trading aggressively and driving the price further up or down on the crash, other players see an increase in traded volume which they interpret as liquidity and start adding to the imbalance in order flow, as they try to reduce their losses during the crash. The careful reader can indeed understand from here why these events are particularly characterised by unbounded traded volume distributions which originate from dynamics of positive feedback loops analogous to those of self-organised criticality \cite{christensen2005complexity}. The dynamics described above induce a positive feedback loop which is bounded by human actions or lack of activity by lower frequency agents, but it is unbounded is size in its underlying nature. This further highlights the need for careful regulation of such events as their potential risk is unbounded.

The reader should be aware that our calibration for jump detection discounts changes in overall volatility and detects anomalously large jumps. These time intervals are not only characterised by large price changes, but by anomalous ones. These originate from avalanche-like negative feedback loops which cause black swan-like trading activity events. 
We highlight that we have also analysed other order flow events such as limit orders and deletions, but trades are the only event we see to be bounded for non-jump intervals and unbounded for jump intervals.
These events are indeed anomalous as particularly trade volumes are carefully calibrated by market participants under normal market conditions in order to minimise impact and detectability of the order flow.

\section{\label{conclusion}Conclusion}

In this work we investigate anomalous price changes in high frequency markets, also known as flash crashes. We analyse the volume distribution over 1 minute time intervals and show that intervals with no anomalous price changes have a tail exponent $>2$ which implies a well defined variance. Intervals with anomalous price changes present a volume distribution with tail exponent $<2$ which is characterised by black swan events and unbounded variance.
Our findings reflect the analyses in the literature suggesting how such anomalous crashes originate from positive feedback loops which involve HFT market players. This is in line with findings in self-organised criticality on natural phenomena such as avalanches.
Correct statistics is crucial for risk modeling, in particular when focusing on extreme events. The unboundedness of the crash volume distribution makes the argument for the use of robust heavy-tailed statistics in volume forecasting and risk modeling in high frequency markets.
Further, the positive feedback loops and the risk of black swan events and market disruptions should prompt market regulators to make sure that risk constraints are being implemented in a fragmented manner across assets, exchanges and players to avoid alignments in trading halts and liquidity withdrawals.

\bibliography{main}

\end{document}